# Formation of free-floating planetary mass objects via circumstellar disk encounters

Zhihao Fu[1,2], Hongping Deng[2]*, Douglas N.C. Lin[3,4], Lucio Mayer[5]

[1]Department of Physics, The University of Hong Kong; Hong Kong, China

[2]Shanghai Astronomical Observatory, Chinese Academy of Sciences; Shanghai 200030, China

[3]Department of Astronomy and Astrophysics, University of California, Santa Cruz; Santa Cruz, CA 95064, USA

[4]Institute for Advanced Study, Tsinghua University; Beijing 100084, China

[5]Department of Astrophysics, University of Zurich; Zurich 8057, Switzerland

*Corresponding author. Email: hpdeng353@shao.ac.cn

**The origin of planetary mass objects (PMOs) wandering in young star clusters remains enigmatic, especially when they come in pairs. They could represent the lowest-mass object formed via molecular cloud collapse or high-mass planets ejected from their host stars. However, neither theory fully accounts for their abundance and multiplicity. Here, we show via hydrodynamic simulations that free-floating PMOs have a unique formation channel via the fragmentation of tidal bridges between encountering circumstellar disks. This process can be highly productive in dense clusters like Trapezium forming metal-poor PMOs with disks. Free-floating multiple PMOs also naturally emerge when neighboring PMOs are caught by their mutual gravity. PMOs may thus form a distinct population that is fundamentally different from stars and planets.**

**Teaser: Free-floating planetary mass objects have their own formation channel, neither star-like nor planet-like.**

## Introduction

Free-floating planetary-mass objects (PMOs) with masses below the deuterium burning limit (*13* Jupiter Mass, $M_J$) are discovered via both direct imaging (*1*,*2*) and microlensing (*3*,*4*). Isolated PMOs that are more massive than Jupiter, lying at the border between stars and planets, are particularly interesting. Such PMOs glowing in infrared were frequently observed in nearby young star clusters by the James Webb Space Telescope (JWST) (*5*-*8*). Are they representatives of the low-mass end of the star formation process or unlucky giant planets exiled by their parent stars?

Stars formed by molecular cloud collapse follow characteristic mass functions. However, in the 10-20 $M_J$ mass range, isolated objects can be excessively abundant, as observed in the Trapezium cluster (*9*), and free-floating PMOs in the Upper Scorpius stellar association are up to 7 times more abundant than the mass functions' prediction (*2*). As a result, an extra formation channel is required to explain the rich PMO population. In addition, free-floating multiple PMOs and candidates were reported with projected separations ranging from several to hundreds of astronomical units (au) (*5*,*7*,*10*-*12*). The high multiplicity (~ 9%) of PMOs in the Trapezium cluster, though needs further confirmation, defies a star-like origin since stellar multiplicity decays with stellar mass, and the by extrapolation wide binary fraction for PMOs should be close to zero (*13*).

Nor are they mature planets ejected after dynamic interactions. PMOs formed as ejected planets would have spatial and kinematic distributions distinct to those of stars, contrasting with the PMOs' distribution in NGC 1333 (*14*). Dynamic simulations of the Trapezium cluster also suggest the concurrent formation of PMOs with stars (*15*), further supported by the prevalence of extended gaseous disks around PMOs (*16*-*18*). To explain free-floating PMOs via planet ejection begs knowledge of the population of wide-orbit giant planets in the first place, which is poorly constrained. Nevertheless, to explain the PMO population in the Trapezium cluster, an unrealistically high occurrence rate of wide-orbit (>100 au) giant planets is required (*19*,*20*) in contrast with observations (*4*,*21*).

## Results

Here, we propose a scaled-down version of filament fragmentation, the modern theory of star formation (*22*), as a possible means to form free-floating PMOs, including multiples. When the mass per unit length (line mass) of filamentary molecular clouds is beyond a critical value, $M_{\rm crit} = \frac{2c_s^2}{G}$ (here $c_s$ is the sound speed and $G$ is the gravitational constant), they fragment on a scale about four times the filament diameter to form dense cores and eventually stars (*23*). The typical width of such a filament is 0.1 pc with a column density of 0.001 g/cm$^2$ (*24*). To fragment into PMOs, a filament 0.001 times thinner (20 au) with a column density around 1000 g/ cm$^2$ is necessary, assuming a comparable temperature. We show with hydrodynamic simulations that such filaments are naturally produced in the tidal bridge connecting two encountering young circumstellar disks. The filament further fragments into PMOs, sometimes forming binary and even triple PMOs (Figure 1).

## From disk encounters to dense filaments and cores

Many young circumstellar disks are prone to instabilities due to the self-gravity of disk gas (*25*,*26*), potentially leading to disk fragmentation and the formation of gaseous planets (*27*,*28*). Circumstellar disks appear even more unstable when perturbed by a stellar or circumstellar disk flyby. These flybys can induce the formation of PMOs in disks that are otherwise stable in isolation (*29*-*32*). However, brown dwarfs instead of PMOs are often formed in these early studies, and PMOs are only marginally resolved if they are not spurious fragments due to poor resolutions (*32*-*34*).

We performed a series of hydrodynamics simulations of circumstellar disk encounters with the Meshless Finite Mass (MFM) scheme (35) at a mass resolution of 0.0001 Jupiter mass, improved by over an order of magnitude compared to the previous studies. In addition, we focus on the extended tidal bridge, which forms most effectively in near coplanar prograde encounters when the disk's body is in quasi-resonance with the flyby orbital motion (*36*,*37*). Informed by observations of the Orion Nebula Cluster (ONC), we construct models of disks that are marginally stable in 100-200 au (Fig. S1) and surround low-mass stars of ~0.3 solar mass (*38*,*39*). These disks are set onto hyperbolic encounter orbits with periapsis distance $r_p = 200 - 500$ au and velocities at infinity $v_\infty = 1 - 5$ km/s (Table S1, Figure 2).

Previous studies show that Keplerian motion is most strongly perturbed when the peak angular velocity of the flyby is about 0.6 times the Keplerian frequency (*37*). For encounters with $v_\infty = 2 - 3$ km/s and $r_p = 300 - 400$ au, the peak orbital angular speed (at periapsis) is right about 0.6 times the disk Keplerian frequency at 100 au causing strong tidal perturbations (Figure 2E). Besides the periapsis, the orbital angular speed is lower than the peak value (Fig. S2), so the quasi-resonance will sweep the whole region beyond 100 au. As a result, the disks each form two grand spiral arms, and the neighboring arms join to form an extended tidal bridge (Figure 1, Movie S1).

The middle part of the tidal bridge contracts into thin filaments with line mass over the critical value for stability (Figure 2F), forming up to 4 cores in one encounter (Figure 2). ~~However~~ In addition, the exact number of compact cores is determined by the length of the filaments and is sensitive to random density fluctuations (*23*), which is barely predictable from encounter parameters. For example, varying $v_\infty$ by only 50 m/s in the rp400v2.65 model can lead to different results of one core, two cores, and no cores, as shown in Table S1.

For high- and low-velocity encounters, the tidal bridge is either stretched too thin or torn apart by the stars, and thus forming isolated cores becomes impossible (Fig. S3). Surprisingly, the Trapezium cluster with an observed velocity dispersion of 2-3 km/s (*41*,*42*) hits the sweet spot for forming isolated cores via disk encounters (Figure 2E).

PMO properties

The dense cores collapse spherically in almost all simulations within ~400 years (Figure 1), leading to prohibitively small simulation time steps in the core center. To follow the motion of isolated cores further, we have to introduce sink particles (Fig. S4) at the center of dense cores (*43,44*). We follow the system until the parent stars are well separated by >2000 au, then determine the free-floating objects' mass (likely overestimated, Fig. S4, S5), boundness, and disk property (Figure 3).

As shown in Figure 2, the coplanar prograde encounters are very productive in forming isolated dense cores. For $r_p = 300 - 400$ au, $v_\infty = 2 - 3$ km/s, almost every encounter (see Table S1) produces a free-floating object (FFO). These FFOs have slow velocities relative to the parent stars and can naturally blend into the cluster (*14*). PMOs preferentially form in filaments with at least two cores featuring competing accretion; otherwise, low-mass brown dwarfs of ~20 $M_J$ are formed. This result aligns with the overabundance of free-floating objects of mass 10-20 $M_J$ seen in the Upper Scorpius stellar association (*2*) and Trapezium cluster (*9*). On the other hand, nonsymmetric encounters involving two different disks tend to form a higher fraction of PMOs than the fiducial symmetric encounters (Table S2, Fig. S5). However, the exact masses of free-floating objects formed via disk encounters can be sensitive to gas thermodynamics, determining the filament's width and fragmentation scale (Fig. S6). PMOs are exclusively produced at high efficiency (up to 7 free-floating PMOs per encounter) in test simulations with local isothermal gas (Table S3). Other physical mechanisms to limit the mass of FFOs to the planetary mass regime may include magnetic fields, which appear efficient in a similar process of disk spiral arm fragmentation (*28*).

Our fiducial simulations formed four binaries among the 33 free-floating objects (Table S1), and their semi-major axes of their orbits are 7-15 au (Table S4). The multiplicity fraction is 13.8%$\pm$ 6.9% (one-sigma uncertainty), comparable to the inferred multiplicity of 9% in the 1 million year (Myr) old Trapezium cluster. The multiplicity fraction is even higher if we focus on PMOs, which preferentially form in a chain of competing cores (Figure 2). Their separations may expand to hundreds of au like those in the Trapezium cluster, and a fraction of them become ionized after interacting with cluster stars for millions of years (*45*). On the other hand, the chaotic isothermal simulations, featuring interactions among a chain of close-packed PMOs, preferentially form multiple PMOs, including hierarchical triple and quadruple systems (Fig. S7, Table S4). The tight binaries in the hierarchical triple PMOs have semi-major axes < 4 au (*10*), while the tertiary PMO is loosely bound and may leave the system (*45*).

Extended disks form around PMOs in our fiducial simulations (Figure 3, Table S1), with low-density gas filling the PMOs' Hill radii up to 200 au (*16*). These disks have a characteristic surface density profile which is nearly flat within 10 au then declines quickly, following a $R^{-2}$ scaling (Fig. S8); the disks show Keplerian rotation within 10 au then transit to sub-Keplerian rotation. On the other hand, disks around multiple PMOs are dynamic and eject mass into the ambient environment at every periapsis crossing of PMOs (Movie S1). In addition, PMOs and their hosts are expected to be metal-poor since they inherit materials in the parent disks' outskirts that are susceptible to dust drift and, thus, are metal-depleted (*46*). The disk may quickly lose most of its mass when it is subject to photoevaporation, as in the Trapezium cluster (*16*), and potentially leave traces as tiny molecular clouds, i.e., globulettes (*47*). As a result, it is uncertain if they can form scaled-down planetary systems in the long term. Nevertheless, the characteristic disk size, steep density profile, and metal-poor nature confronted with observations (*16-18*) may help to tell if PMOs are formed via disk encounters.

We further tested the effect of misaligned disks to assess PMO formation efficiency in general encounters. Slightly misaligned encounters evolve similarly to coplanar encounters, as shown in Figure 1, but form less dense tidal bridges. When the mutual inclination of disks is smaller than the disk opening angle (about 5 degrees here), free-floating objects can form via tidal bridge collapse (Table S5). However, this should not be regarded as a solid criterion since encounters with more massive disks can form more critical filaments and still fragment into free-floating objects.

## Discussion

In the Trapezium cluster, we can estimate the close encounters rate for $r_p <$500 au, as $\sigma_v n \pi b^2$, here we take a stellar density $n = 5 \times 10^4$ $pc^{-3}$, velocity dispersion $\sigma_v = 2.5$ km/s (*41*) and $b$ is the corresponding impact parameter. We find every star experiences 3.6 encounters on average within 1 Myr, i.e., the lifetime of Trapezium. We estimate ~10% of the encounters to be nearly coplanar (relative angle < 5 degrees), informed by the distribution of relative angles for binary disks in star cluster simulations (*48, See Supplementary Text*). The highly efficient PMO production channel via encounters (Figure 2, Fig. S6) can therefore explain the hundreds of PMO candidates (540 over 3500 stars) observed in the Trapezium cluster. In addition, PMOs so formed have a high initial multiplicity fraction that can account for the present-day PMO binary fraction in Trapezium even after some ionization.

The Upper Scorpius Association has the next largest known population of free-floating PMOs, second to the Trapezium cluster. It also features a velocity dispersion (*49*) amenable for PMO formation via disk encounters (Figure 2). Nevertheless, the IC 348 and NGC 1333 clusters have smaller velocity dispersions <1 km/s (*50,51*), affecting the encounter rates and PMO formation efficiency and, leading to a smaller population of PMOs than that in Trapezium. This outcome is particularly true for IC 348, which has a low stellar density. Future studies of various young clusters (*42*) can further constrain the population of PMOs.

## Materials and Methods

### Numerical methods

We employed a Godunov-type Lagrangian method, the Meshless Finite Mass (MFM) scheme (*35*), to solve the self-gravitating hydrodynamics in two interacting massive circumstellar disks. Specifically, the simulations are performed with the public hydrodynamic code GIZMO, where gas self-gravity is efficiently calculated via a Tree algorithm with adaptive gravitational softening (*35,52,53*). The MFM method has been successfully deployed in modeling star cluster formation (*44*) and gravitationally unstable circumstellar disks focusing on disk fragmentation (*28,54*). Notably, MFM can avoid artificial fragmentation even in poor resolution, producing consistent results across a wide range of resolutions (*44,54*). In contrast, artificial fragmentation often occurs in low-resolution SPH simulations (*32-34*). Thanks to MFM's Lagrangian nature and good conservation property, we managed to cover a size range of 0.1 au (in disks around PMOs, Figure 3) to >2000 au in these encounter simulations (Figure 1).

For simplicity, an idealized barotropic gas equation of state (EOS) (*55*), calibrated to radiation-hydrodynamic simulations of the collapse of a protostellar core of 1 solar mass (*56*), is used to allow an extensive suite of high-resolution simulations. However, this EOS likely underestimates the cooling efficiency (*57*) because 1) the PMO cores are much less massive than protostellar cores, and 2) both the cores and the tidal bridge hosting them should be metal-poor and thus have low opacity (*46*). Thus, we also provide isothermal simulations for comparison. We note that the collapse of isothermal filaments cannot be halted by the gas pressure (*23*), so the filament width keeps decreasing until small-scale supersonic turbulence provides support at a certain point. Thus, the filament width and core separation are quite uncertain. More detailed radiative hydrodynamics simulations are desirable in the future and are expected to give results between our barotropic and isothermal simulations.

As the dense cores collapse (Figure 1), the time steps within the cores become very small, preventing long-term simulations, so sink particles are introduced at their centers per the criteria of Ref *44*. We use a simple accretion criterion (*43*) for all sink particles accreting material within a sink radius of 0.5 au (10 au for the stars hosting the disks). The mass and momentum of accreted materials are added to the sink particle to ensure mass and momentum conservation. This approach inevitably exaggerates the accretion of the sink particles. As a result, their masses should be regarded as the upper limits of the objects' actual masses.

Initial Conditions

We use observationally informed parameters for the circumstellar disks, which closely represent conditions in the Trapezium cluster, where many PMOs have been observed. We consider young circumstellar disks given the young age of the Trapezium cluster of about 1 million years (Myr). They are relatively massive and likely hover around the edge of gravitational instability (*26,58,59, Supplementary Text*). The disks are assumed to follow a self-similar surface density profile (*60*),

$$\Sigma(R) = \Sigma_0 \left(\frac{R}{R_c}\right)^{-\gamma} \exp\left[-\left(\frac{R}{R_c}\right)^{2-\gamma}\right].$$

Here, we set $\gamma = 0.9, R_c = 200$ au, and the disk extends from 20 au to 600 au (*59,61*). The disk mass is 0.18 (0.2) solar mass surrounding a central star of 0.3 (0.33) solar mass (*39*). They are named by the host mass and disk mass as $m_s0.3m_d0.18$ or $m_s0.33m_d0.2$ (Fig. S1).
Each disk is represented by 1.8 M or 2 M fluid particles required to resolve the disk's vertical structure and avoid spurious fragmentation (*32-34*). Equivalently, a 10 Jupiter mass PMO is resolved by 100,000 particles. We also considered low-resolution models with particles two times more massive. We realized the disk density structure via rejection sampling following Ref *34*. To reduce the particle noise, we relaxed the initial condition for 5000 years by damping the vertical and radial motion at every timestep to reach an equilibrium state.

These disks are close to gravitational instability with an initial Toomre Q parameter (*62*) of about 2 (Fig. S1) between 100 au and 200 au (*25,26*). We also considered for tests an isothermal disk with a temperature profile proportional to $R^{-0.6}$ where the temperature at 1 au is assumed to be 300 K and a minimum temperature of 10 K is applied at the outer part (*34*). Over 90% of the disk mass resides within 400 au, about the upper limit of the observed disk size in the Trapezium cluster (*38*). However, we note that the encounters themselves will notably truncate the disks, as shown in Fig. S2. In addition, long-term photoevaporation likely shrinks the disks in size further (*63*), so we believe the adopted disk size is reasonable.

Simulation Setup and Analysis

We place isolated circumstellar disks on prescribed hyperbolic trajectories with an initial separation of 2000 au. Here, we focus on coplanar encounters with prograde spins for both disks, often leading to long tidal bridges critical for PMO formation. However, previous studies focused on noncoplanar cases (*31*) or encounters with retrograde disk spin (*32*).

These hyperbolic orbits are characterized by their periapsis distance ($r_p$) and velocity at infinity ($v_\infty$). Our fiducial simulations cover $r_p$~ 200-500 au and $v_\infty$~1-5 km/s for encounters involving two $m_s$0.3$m_d$0.18 disk models (Fig. S1), and we summarize the results in Table S1. These simulations illuminate the critical physics of PMO formation and identify the relevant parameter space. We then studied encounters involving two different barotropic disks (see Fig. S1) but focused on $r_p$~ 300-400 au and $v_\infty$~2-3 km/s, i.e., the PMO productive region of Table S1. These nonsymmetric encounters are summarized in Table S2. We further tested the effect of gas thermodynamics by employing the isothermal disk model in Fig. S1 and present the results in Table S3. Finally, we explored a few cases of noncoplanar encounters in Table S5 and performed simulations at a lower resolution (employing ~2M particles) to study the resolution effect in Table S6.

We simulate the encounter until the stars are separated by >2000 au again or up to 7500 years (only for low-velocity encounters unable to produce free-floating objects). The duration of the simulation is primarily limited by the computational cost. Each encounter simulation takes up about 20,000 CPU hours on the ShanHe supercomputer of the Chinese National Supercomputing Center, Jinan. The stars follow the hyperbolic trajectory well before the encounter (Fig. S2), while the orbit tightens due to later tidal interactions (*64*). After the encounter, $v_\infty$ typically declines by 0.1- 0.3 km/s for encounters with $r_p \geq 300$ au and can decline by 0.6 km/s for a close encounter with $r_p = 200$ au. Near the end of the simulations, the tidal bridge is substantially dispersed with a column density below 1 g/cm$^2$ (Figure 1). We then check if the formed objects are unbound to the parent stars to become free-floating objects (FFOs). We also check if FFOs are bound to each other to form multiples. Disk masses for single FFOs are calculated by weighting gas within its Hill radius with respect to the nearest star.

**Acknowledgments:** We acknowledge the stimulating discussions with Fabo Feng, Su Wang, Dong Lai, Andrew Winter, Xuening Bai, Greg Herczeg, Wei Zhu, Xuesong Wang, and Yihan Wang. We acknowledge the anonymous referees' kind suggestions and constructive comments, which greatly improved the paper. The simulations were performed at the National Supercomputing Center, Jinan, and the prompt support from their technicians is highly appreciated.

**Funding:** HPD is supported by a talent program from the Chinese Academy of Sciences.

**Author contributions:**

Conceptualization: HPD

Methodology: HPD

Investigation: ZHF, HPD

Visualization: ZHF, HPD

Funding acquisition: HPD

Project administration: HPD

Supervision: HPD

Writing – original draft: HPD

Writing – review & editing: ZHF, HPD, DNCL, LM

**Competing interests:** All authors declare no competing interests.

**Data and materials availability:** All data needed to evaluate the conclusions in the paper are present in the paper and the Supplementary Materials. The GIZMO code for the hydrodynamic simulations is publicly available at http://www.tapir.caltech.edu/~phopkins/Site/GIZMO.html. The simulation data presented in the figures and supplementary figures are available at https://doi.org/10.5281/zenodo.14403644, and the visualization made use of the SPLASH visualization tool at https://github.com/danieljprice/splash.

**Supplementary Materials**

This PDF file includes

Supplementary Text

Figs. S1 to S8

Tables S1 to S6

Movies S1 to S2

References

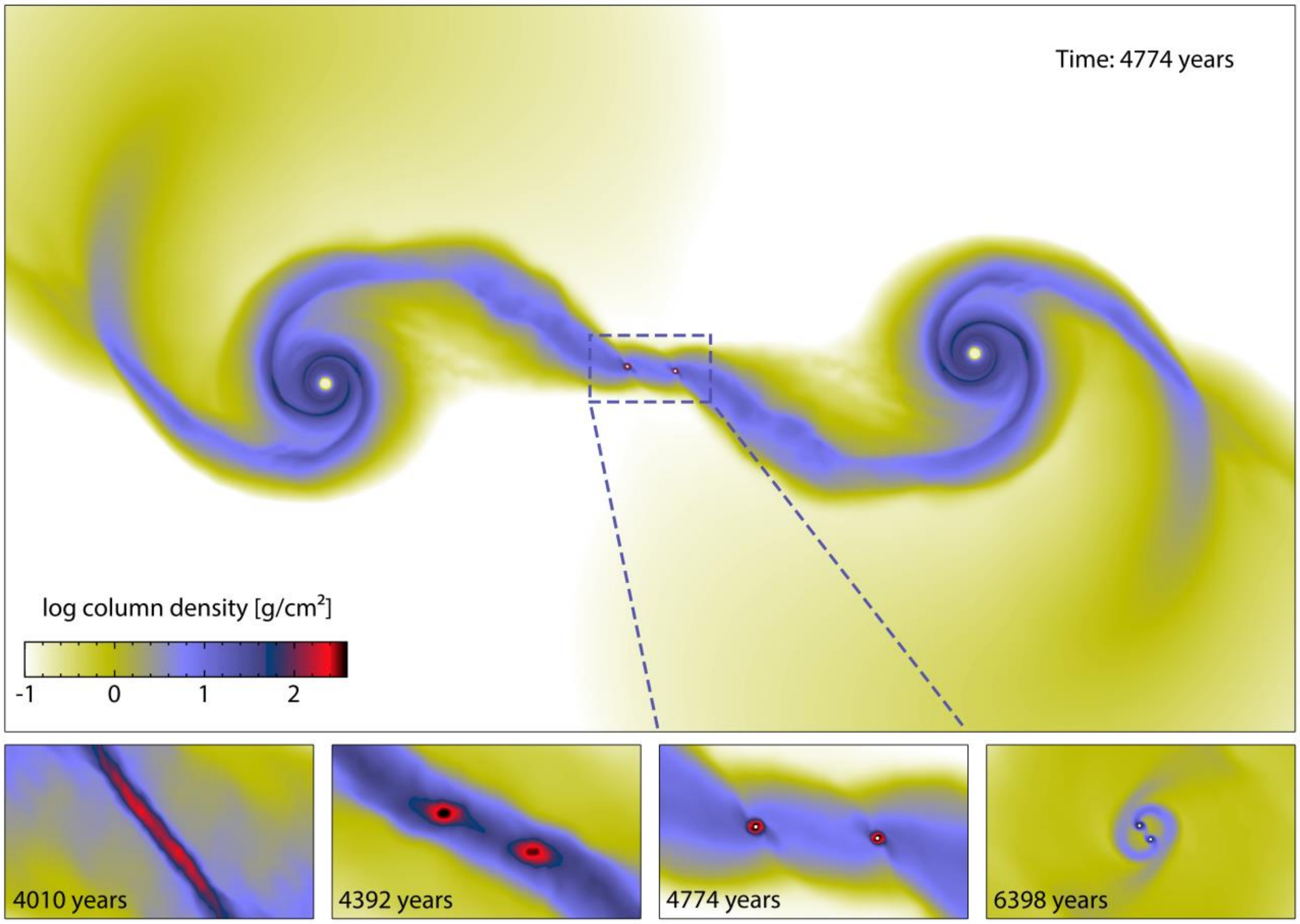

**Figure. 1. The formation of binary PMOs via circumstellar disk encounters.** The upper panel shows the density map (2000 au × 1120 au) in a logarithmic scale for an example of a simulation (model rp400v2.65 of Table S1). The lower panels enlarge a region of 200 au × 112 au to show the evolution of the binary PMOs, sink particles in white with exaggerated radii, and how they emerge within the dense filament created by the encounter (see Movie S1).

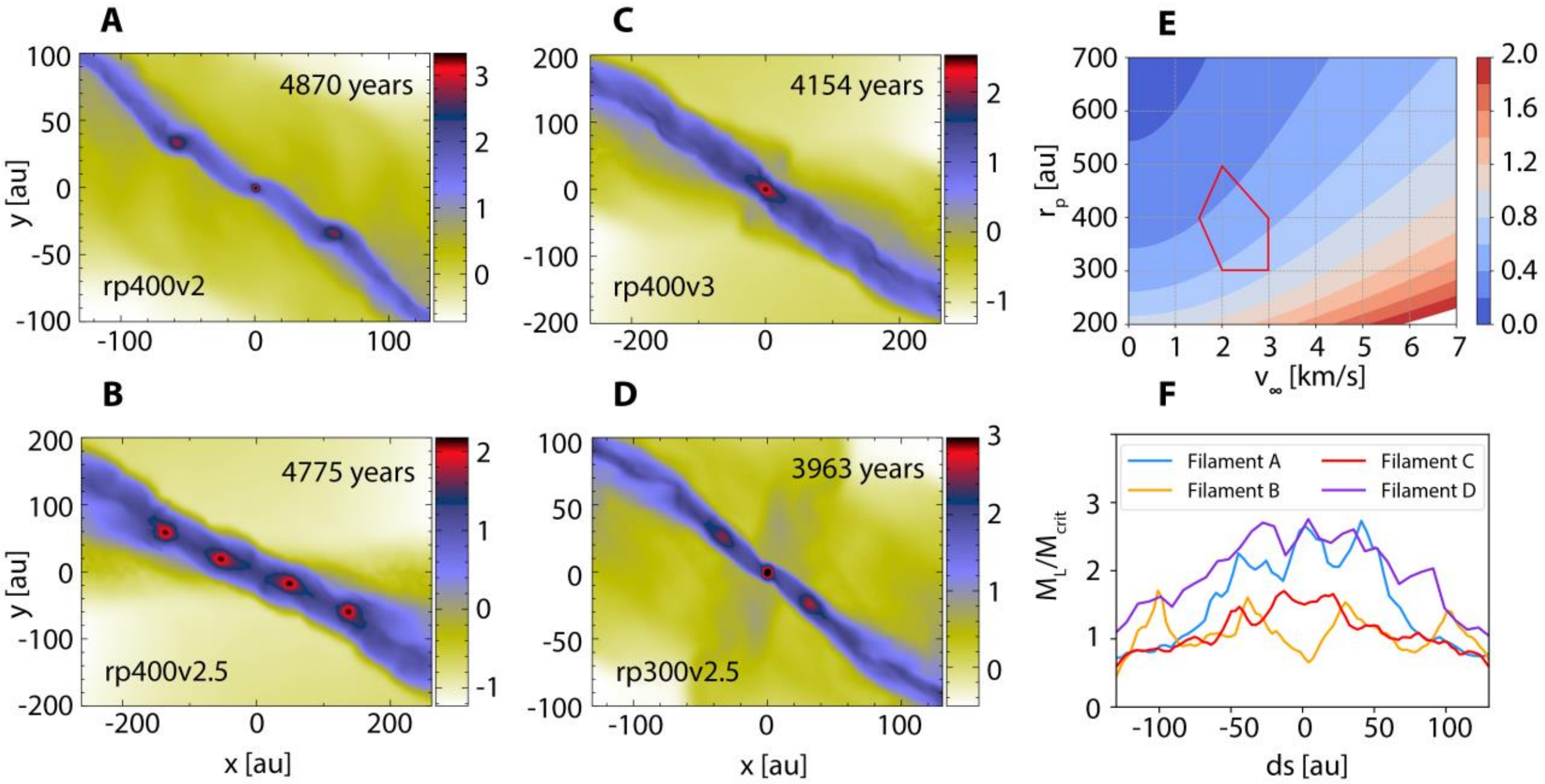

**Figure 2. The formation of compact cores within dense filaments under different encounter conditions.** Panels (A) to (D) show the central region gas column density in g/cm$^2$ in a logarithmic scale for representative simulations in Table S1. They possess a series of dense cores comparable to the Figure 1 insert panel at 4392 years. Panel (E) identifies the encounter parameter space that forms free-floating objects via filament collapse with red curves (Table S1); the background contours show the peak angular speed of different flybys divided by the Keplerian frequency at 100 au. Panel (F) shows the line mass of precursors of the (A) to (D) filaments (~400 years earlier; see also Figure 1) normalized to the critical line mass for stability (see main text).

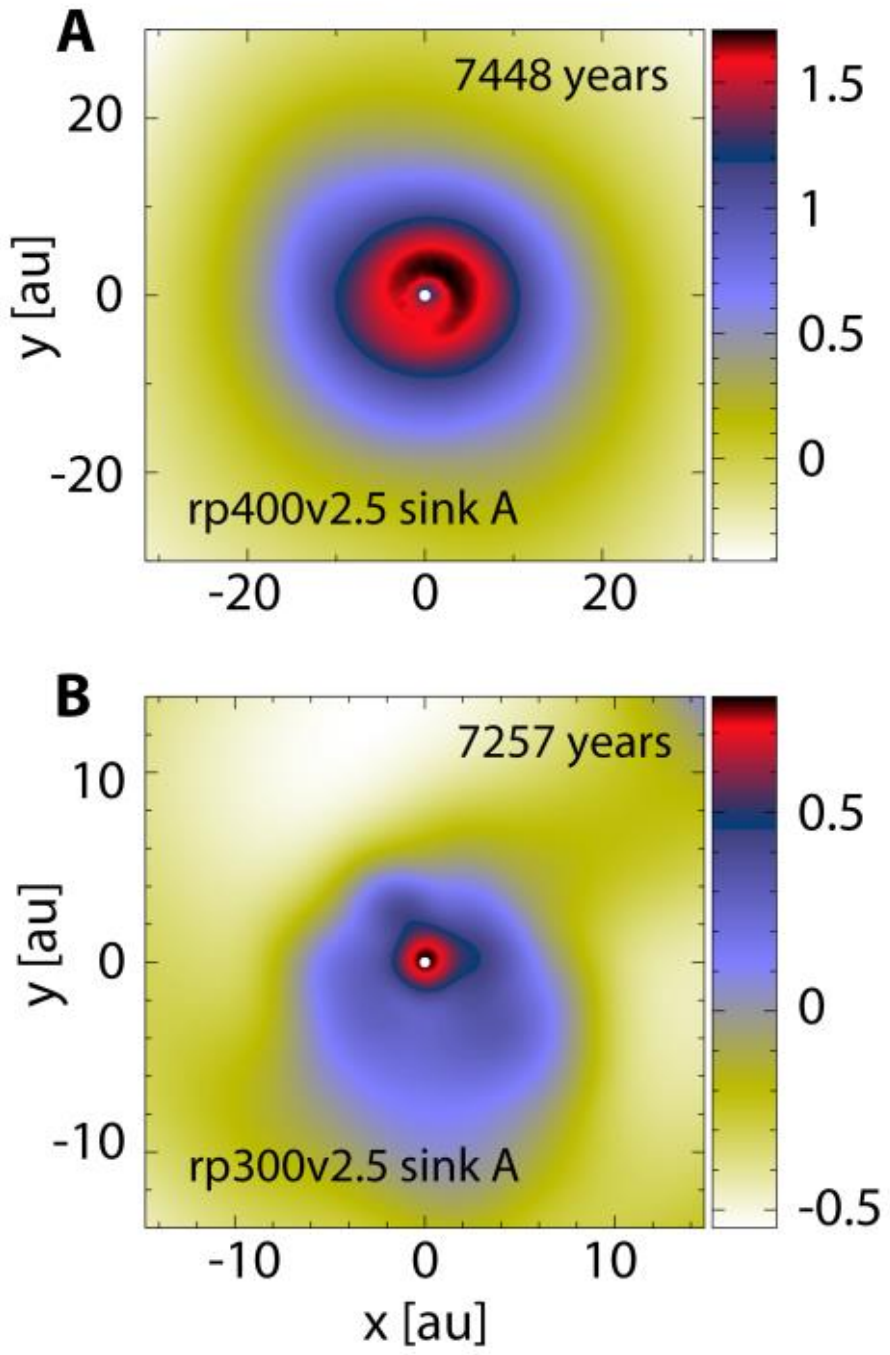

**Figure 3. Disks around free-floating PMOs.** Panels (A) and (B) show similar column density maps like Figure 1, centered on free-floating PMOs at the end of two simulations (Table S1). The PMOs can have extended disks, and the PMO disk in panel (B) is lopsided because of perturbations from the other PMO (to the upper right, not shown here) in this binary system (see also Figure 1).

# Supplementary Materials for

## Formation of free-floating planetary mass objects via circumstellar disk encounters

Zhihao Fu *et al.*
*Corresponding author. Email: hpdeng353@shao.ac.cn

**This PDF file includes:**
Supplementary Text
Figs. S1 to S8
Tables S1 to S6
Movies S1 to S2
References

**Supplementary Text**

Estimations for the probability of near coplanar encounters

The statistics of the mutual inclination between encountering disks are poorly constrained by both observations and theory. However, the pairs of disks encountering each other should have spins correlated instead of randomly drawn. The reason is twofold: 1) close-by stars/disks forming in the same molecular cloud filaments have a higher probability of encountering their close siblings than a random star farther away (assuming a stellar velocity dispersion of 2 km/s like in the Trapezium cluster, the distance a star can travel in 1 myr is only 2 pc comparing to the cluster size of ~400 pc); 2) stars/disks forming in the same molecular cloud filaments mostly have spin perpendicular to the initial filament axis (*64*), i.e., they have almost aligned spins.

Although the distribution of mutual angle between encountering disks is unknown, the mutual angle between binary disks can be a reasonable approximation because recent observations show that the relative angle between binary disks is independent of the projected separation (*65*). Notably, the orientation of binary disks is correlated instead of randomly drawn (*66*); otherwise, we would not expect the two near coplanar binary disks among the seven binary disks of Ref *65*.

To roughly estimate the probability of near coplanar encounters, we turn to the star cluster formation simulation by Ref *67*. We utilize the distribution of relative angles between binary disks (Fig. 19 of Ref *47*) to estimate a probability of near coplanar encounters of ~10%. The PMO production rate by encounters can vary due to disk property, thermodynamics, and encounter geometry (table S1-S3, table S5). In general, a coplanar encounter probability of a few percent agrees with the PMO fraction in Trapezium.

On the lifetime of marginally gravitationally stable disks

Our disk models are marginally gravitationally stable (fig. S1), which is the likely state of early disks subject to infall maintaining a balance between material loading and transport (*26,68*). By early disks, we refer to class 0 and class I disks, covering the first 1 myr evolution. For example, the class I HL tau disk is fed by infall and streamers (*69*) and is still close to gravitational instability (*70*). These young disks are massive with disk-to-star mass ratios > 0.1 (*70,71*).

On a population level, recent studies suggest that some disks of several myr old may still be massive enough and are at the cusp of gravitational instability (*58*). As a prominent example, the 2.5–4.4-myr-old AB Aurigae can have a disk of up to a third of the stellar mass (*57*). In summary, Class II marginally gravitationally stable disks are probably no more than 10-20% of the population, but Class I objects are susceptible to gravitational instability for a few hundred thousand years (*68*), and streamers may extend that state beyond 1 myr. As a result, it is safe to assume that the encounters that happened within the 1 myr old Trapezium cluster involve disks marginally gravitationally stable.

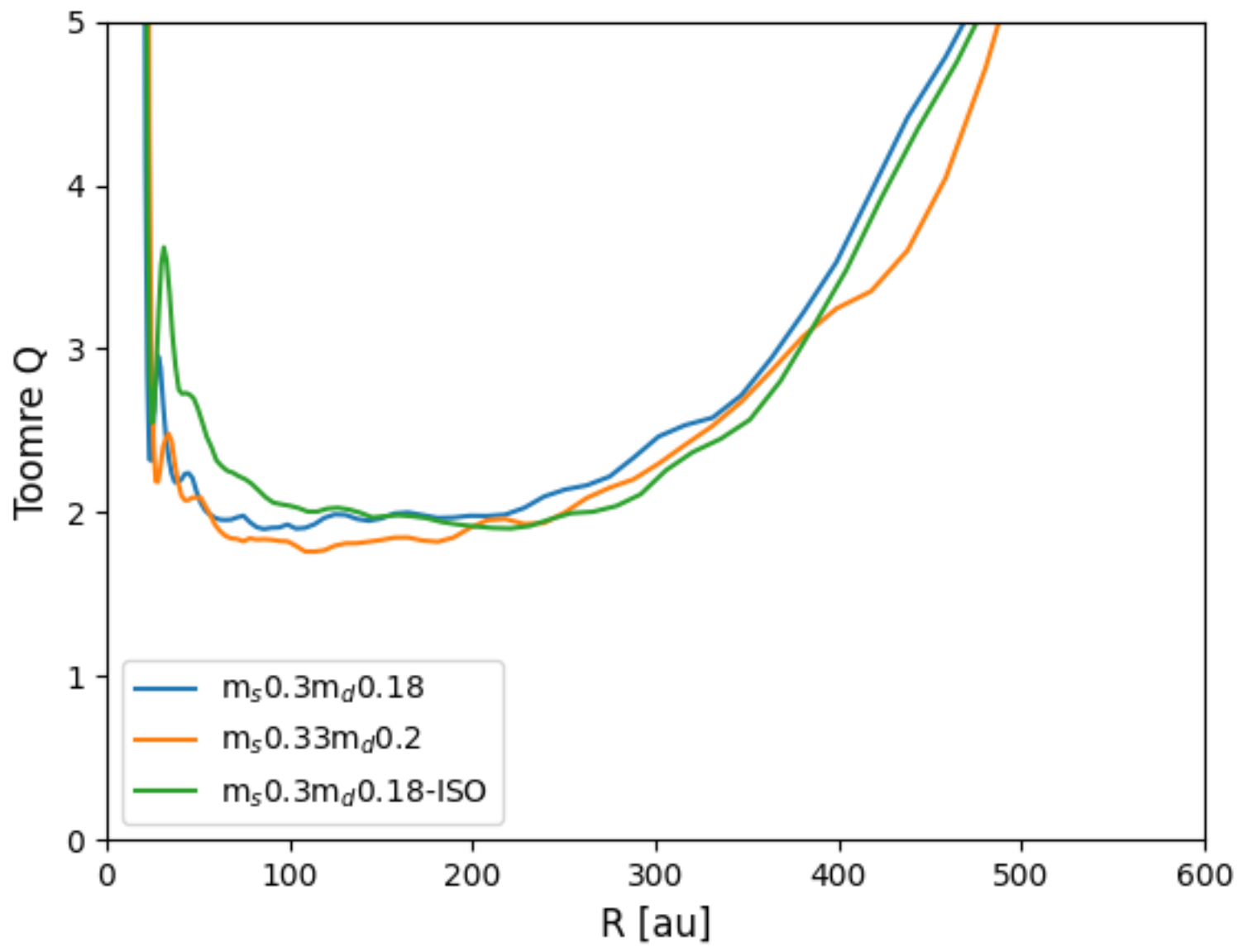

**Fig. S1.**

**The Toomre Q profiles for isolated disk models with different mass and EOS.** The disks are marginally gravitationally stable in the 100-200 au region. The legend indicates the host mass ($m_s$) and the disk mass ($m_d$), and the suffix "ISO" indicates an isothermal model.

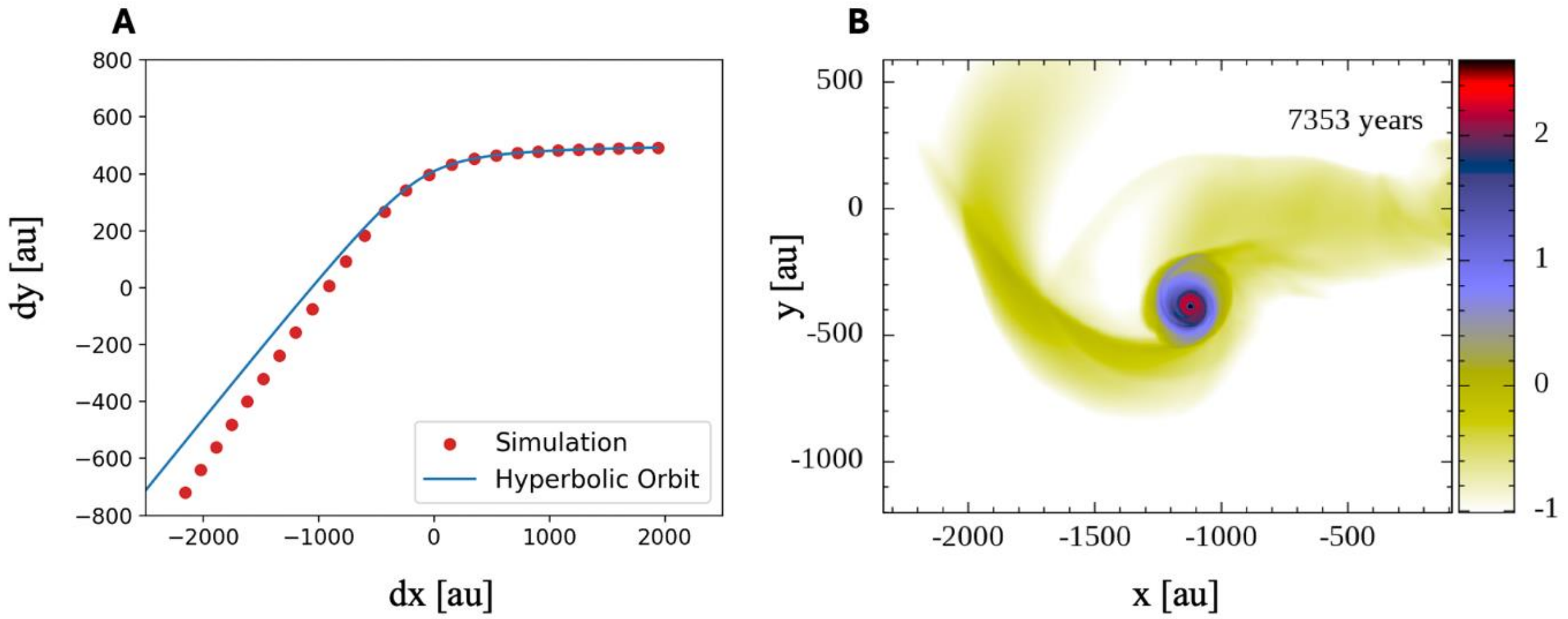

**Fig. S2.**

**The relative trajectory of the stars (A) and the disk morphology at the end of the Fig. 1 simulation (B).** The time interval between two consecutive simulation snapshots in (A) is 286 years. The post-encounter disk possesses two streams of material extending beyond 1000 au which should be distinguished from infall.

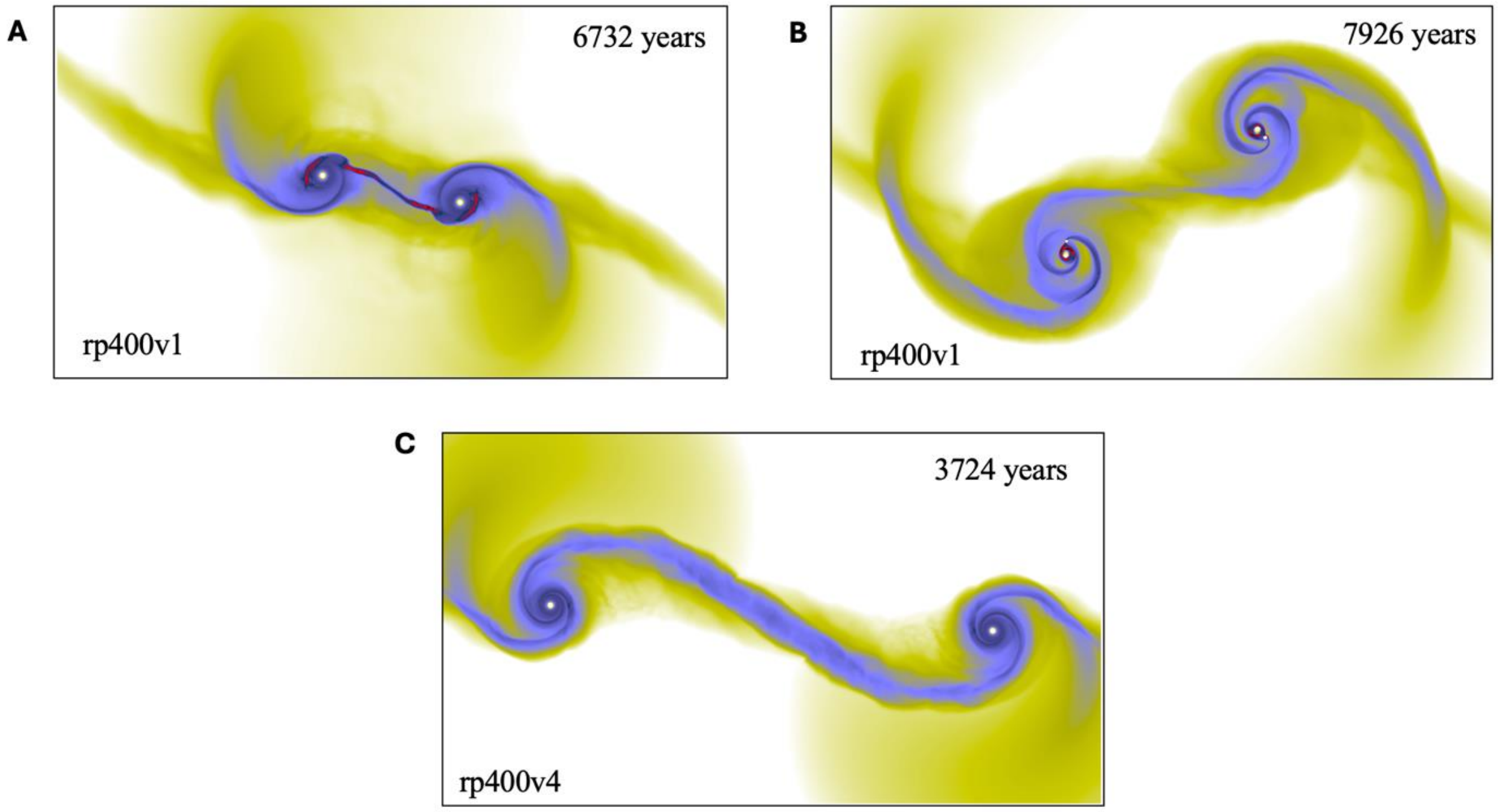

**Fig. S3.**

**The formation of bound companions in a slow encounter (A, B) and the diffusive tidal bridge in a fast encounter (C).** Each panel shows the logarithm of the surface density in g/cm$^2$ covering a region of 2000 au × 1120 au centered on the origin (like Fig. 1). In the low-speed encounter with $v_\infty = 1$ km/s, the tidal bridge is short. It is torn apart by the nearby star and eventually collides with the other spiral arms to form bound companions. However, the long tidal bridge in fast encounters is quickly dispersed due to fast stretching.

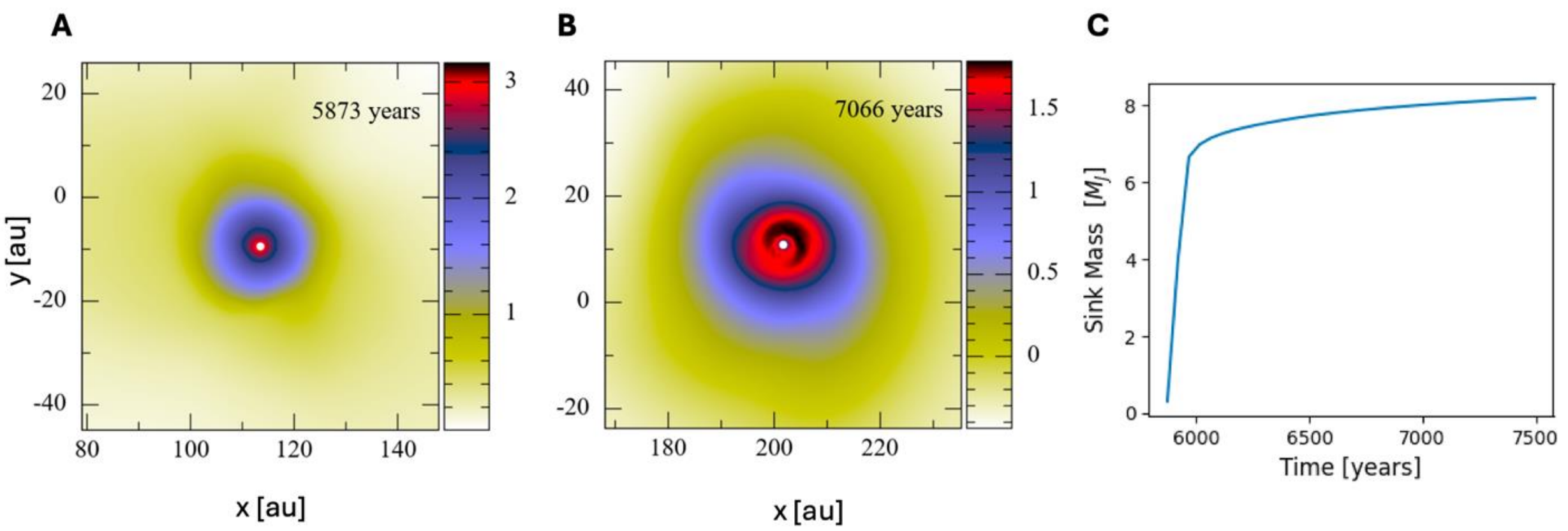

**Fig. S4**

**The evolution of sink A in the rp400v2.5 model of table S1 (see also Fig. 3A).** Panels (A) and (B) show the column density map in g/cm$^2$ in a logarithmic scale; the sink radius is 0.5 au and is not shown to scale in the plots. Panel (C) shows the growth of the sink particle mass where the linear growth stage can be regarded as the spherical collapse phase. Later, sink particles create regions of effective vacuum in the cores of PMOs, leading to exaggerated accretion so that their masses are upper limits of the PMOs' true mass. Increasing the resolution can better resolve the flow around PMOs, resulting in less numerical accretion and lower sink mass (*29*, fig. S5).

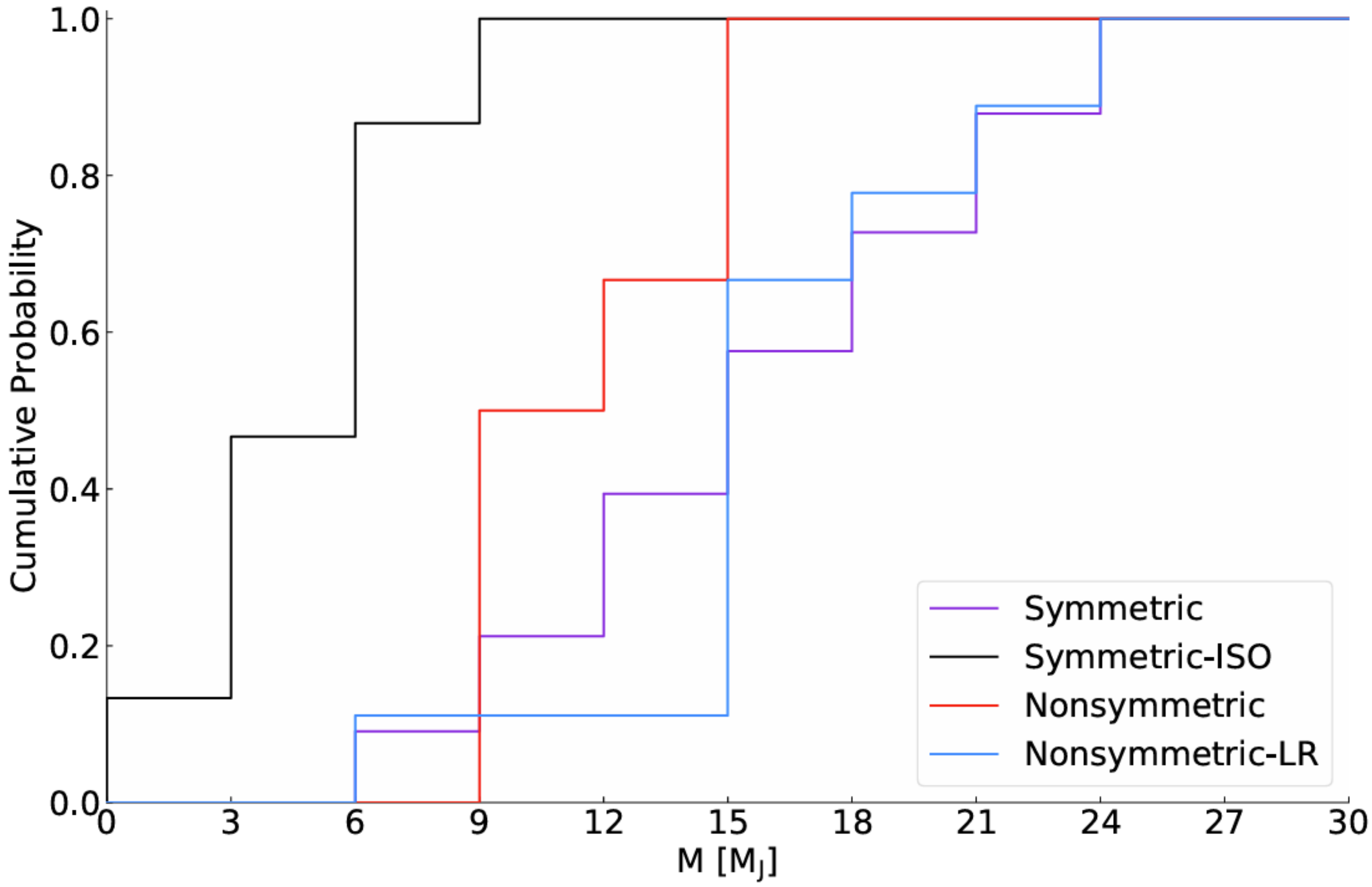

**Fig. S5.**

**Cumulative distribution function of the free-floating objects' (FFOs) mass in symmetric encounters with two identical disks (table S1), symmetric encounter with isothermal disks (table S3), nonsymmetric encounters with two different disks at the fiducial resolution (table S2) and a low resolution (table S6).** Nonsymmetric encounters tend to form more PMOs than symmetric encounters, while isothermal simulations form PMOs exclusively. We expect a larger PMO fraction in the barotropic simulations if the resolution is even higher, given the trend in the resolution test (see also discussion in fig. S4). However, the isothermal test cases featuring almost spherical collapse with minor disks (table S3) likely possess the lowest mass PMOs that can form via disk encounters.

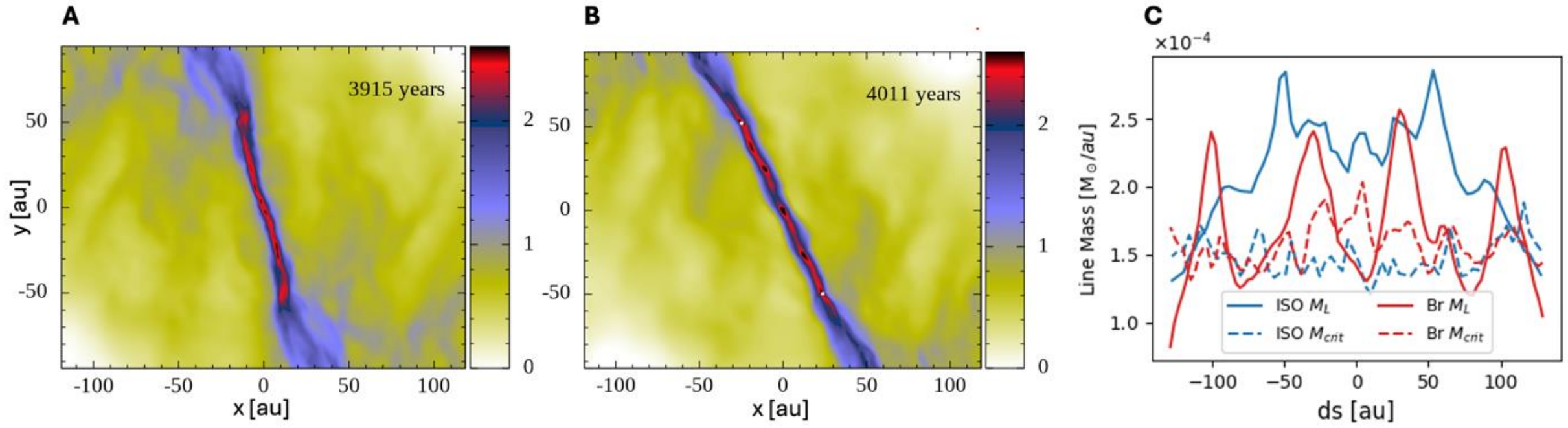

**Fig. S6**

**The filaments in the barotropic (Br) simulation rp400v2.5 (panel A, see table S1) and isothermal (ISO) simulation ir400v2.5 (panel B, see table S3).** Panel (C) compares the filaments line mass (solid lines) to the critical line mass for stability (dashed lines). The filament in the isothermal simulation is more conducive to fragmentation and eventually forms 6 closely packed sink particles (see Movie S2).

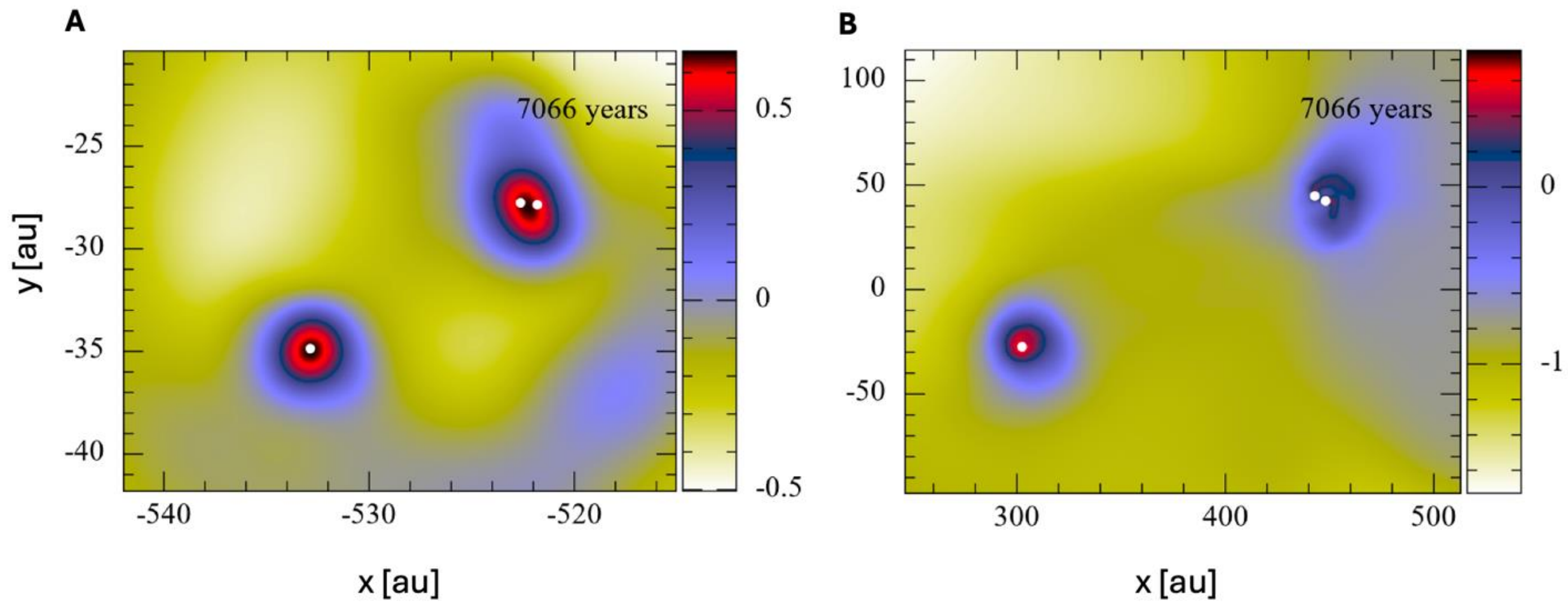

**Fig. S7**

**Two free-floating PMO triples formed in the isothermal simulation ir400v25 of table S3 (see Movie S2).** The gas column density is low, and the tight binaries in panels (A) and (B) are not expected to merge, while the loosely bound third PMO in panel (B) may be ionized due to dynamical interactions with other stars (*44*).

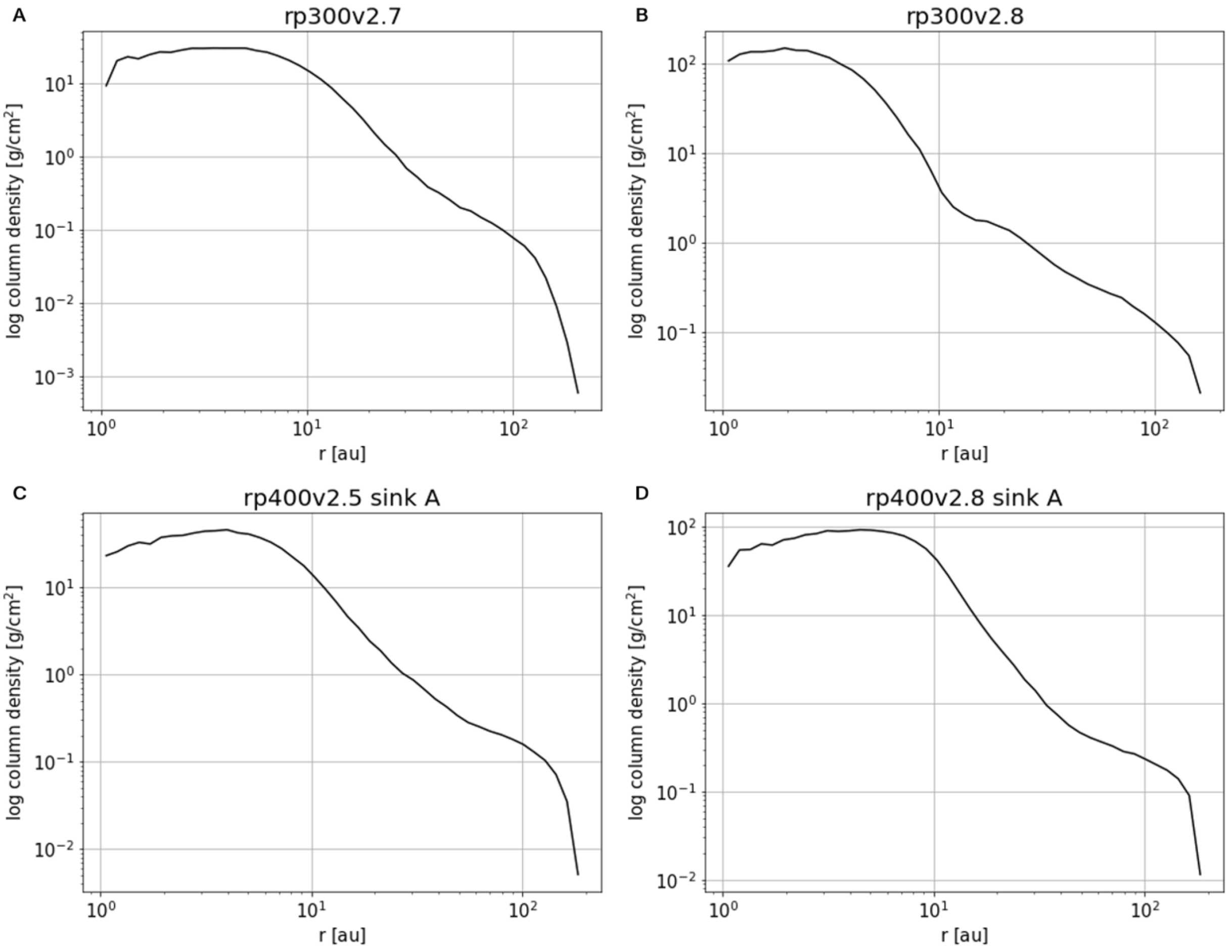

**Fig. S8**

**Panel (A) to (D), surface density profiles of disks around several free-floating single PMOs in table S1.** They have similar profiles that feature a flat region followed by fast decay, which is useful for verifying the PMO formation theory via disk encounters.

**Table S1.**

**The fiducial encounter simulations of two 0.3 solar mass stars, each hosting a 0.18 solar mass disk (fig. S1, $m_s0.3m_d0.18$ model).** The simulations are labeled by the encounter orbits' periapsis distance (in au) and velocity at infinity (in km/s). The number of bound companions (BCs) and free-floating objects (FFOs), including binaries are listed; FFOs' mass is given as the sink particle mass (in Jupiter mass, $M_J$), and all materials within their Hill radii are regarded as their disk. However, it is difficult to define circumsingle disks around multiple FFOs, so we temporarily ignore them. For example, material around single PMOs is substantially ejected during the periapsis crossing of binary PMOs in the rp400v2.65 model (Movie S1).

| Case | Number of BCs | Number of FFOs (binaries) | A Sink mass (disk mass) for FFOs | B Sink mass (disk mass) for FFOs |
|---|---|---|---|---|
| rp200v1 | 6 | 0 | | |
| rp200v2 | 3 | 0 | | |
| rp200v3 | 4 | 0 | | |
| rp200v4 | 0 | 0 | | |
| rp200v5 | 0 | 0 | | |
| rp300v1 | 4 | 0 | | |
| rp300v1.5 | 5 | 0 | | |
| rp300v2 | 0 | 1 | 26.58 (11.55) | |
| rp300v2.1 | 0 | 1 | 21.91 (8.66) | |
| rp300v2.2 | 2 | 0 | | |
| rp300v2.3 | 0 | 1 | 23.48 (8.47) | |
| rp300v2.4 | 0 | 1 | 26.46 (15.69) | |
| rp300v2.5 | 1 | 2 (1) | 10.3 | 14.64 |
| rp300v2.6 | 2 | 0 | | |
| rp300v2.7 | 0 | 1 | 11.33 (2.71) | |
| rp300v2.8 | 2 | 1 | 7.21 (3.09) | |
| rp300v2.9 | 0 | 1 | 22.27 (16.65) | |
| rp300v3 | 0 | 1 | 21.12 (11.7) | |
| rp300v3.5 | 0 | 0 | | |
| rp300v4 | 0 | 0 | | |
| rp300v5 | 0 | 0 | | |
| rp350v2.2 | 2 | 1 | 19.24 (6.57) | |
| rp350v2.3 | 0 | 1 | 25.73 (16.29) | |
| rp350v2.4 | 0 | 2 (1) | 16.29 | 16.57 |
| rp350v2.5 | 0 | 1 | 20.17 (14.39) | |
| rp400v1 | 3 | 0 | | |
| rp400v1.5 | 2 | 1 | 14.34 (5.03) | |
| rp400v2 | 2 | 1 | 15.66 (3.56) | |
| rp400v2.1 | 0 | 1 | 18.79 (12.76) | |
| rp400v2.2 | 0 | 2 (1) | 13.54 | 14.85 |
| rp400v2.3 | 0 | 2 | 17.99 (5.4) | 18.03 (5.72) |
| rp400v2.4 | 0 | 1 | 21.26 (16.64) | |
| rp400v2.5 | 0 | 2 | 8.02 (3.35) | 8.19 (3.38) |
| rp400v2.6 | 0 | 1 | 17.07 (13.1) | |
| rp400v2.65 | 0 | 2 (1) | 12.8 | 13.24 |
| rp400v2.7 | 0 | 0 | | |
| rp400v2.8 | 0 | 2 | 11.39 (7.31) | 11.6 (7.08) |
| rp400v2.9 | 0 | 1 | 15.4 (17.58) | |
| rp400v3 | 0 | 1 | 20.48 (10.47) | |
| rp400v3.5 | 0 | 0 | | |
| rp400v4 | 0 | 0 | | |
| rp500v1 | 0 | 0 | | |
| rp500v2 | 0 | 1 | 24.03 (13.33) | |
| rp500v3 | 0 | 0 | | |
| rp500v4 | 0 | 0 | | |
| rp500v5 | 0 | 0 | | |
| rp600v2 | 0 | 0 | | |

**Table S2.**

**Encounters between two different stars of 0.3 solar mass and 0.33 solar mass hosting a disk of 0.18 solar mass and 0.2 solar mass (fig. S1), respectively.** The nonsymmetric encounters are labeled by the encounter orbits' periapsis distance and velocity at infinity like table S1.

| Case | Number of BCs | Number of FFOs (binaries) | Sink mass (disk mass) for FFOs |
|---|---|---|---|
| nr300v2.1 | 4 | 0 | |
| nr300v2.2 | 1 | 0 | |
| nr300v2.3 | 2 | 0 | |
| nr300v2.4 | 4 | 0 | |
| nr300v2.5 | 1 | 1 | 11.02 (3.07) |
| nr300v2.6 | 2 | 1 | 12.85 (4.22) |
| nr300v2.7 | 1 | 1 | 12.77 (5.79) |
| nr300v2.8 | 2 | 1 | 15.94 (4.21) |
| nr300v2.9 | 0 | 1 | 10.35 (3.88) |
| nr300v3.0 | 1 | 1 | 9.91 (4.83) |
| nr400v2.1 | 1 | 1 | 10.86 (2.56) |
| nr400v2.2 | 1 | 0 | |
| nr400v2.3 | 1 | 0 | |
| nr400v2.4 | 1 | 1 | 11.67 (5.34) |
| nr400v2.5 | 1 | 0 | |
| nr400v2.6 | 0 | 1 | 16.75 (11.64) |
| nr400v2.7 | 1 | 0 | |
| nr400v2.8 | 0 | 1 | 17.94 (6.72) |
| nr400v2.9 | 0 | 1 | 10.66 (5.54) |
| nr400v3.0 | 0 | 1 | 15.19 (7.61) |

**Table S3.**

**Test simulations with isothermal EOS involving the $m_s$0.3$m_d$0.18-ISO disk in fig. S1, to be compared with table S1.** Hierarchical multiple PMOs, including triple (T) and quadruple (Q), are formed due to the interaction of closely packed dense cores (fig. S6, S7).

| Case | Number of BCs | Number of FFPs (Multiples) | Sink mass (disk mass) for FFPs | Sink mass (disk mass) for FFPs | Sink mass (disk mass) for FFPs |
|---|---|---|---|---|---|
| ir400v2.5 | 5 | 6 (2T) | 2.99 | 5.76 | 6.38 |
| | | | 6.96 | 9.58 | 10.58 |
| ir400v2.6 | 2 | 7 (1Q) | 0.96 (0) | 3.2 (0.1) | 3.61 |
| | | | 3.67 (0) | 4.43 | 7.14 |
| | | | 7.43 | | |
| ir400v3.0 | 0 | 2 | 6.31 (3.47) | 6.97 (3.96) | |

**Table S4.**

**Detailed information for multiple FFOs.** We note that in the fiducial simulations of table S1, FFOs are often slightly above 13 $M_J$ but the sink particle mass is likely an overestimation (fig. S4, S5).

| Case | $m_1$ $[M_J]$ | $m_2$ $[M_J]$ | $a_2$ [au] | $e_2$ | $m_3$ $[M_J]$ | $a_3$ [au] | $e_3$ | Note |
|---|---|---|---|---|---|---|---|---|
| rp300v2.5 | 10.3 | 14.64 | 14.7 | 0.65 | | | | Binary |
| rp350v2.4 | 16.29 | 16.57 | 8.1 | 0.27 | | | | Binary |
| rp400v2.2 | 13.54 | 14.85 | 7.68 | 0.27 | | | | Binary |
| rp400v2.65 | 12.8 | 13.24 | 15.18 | 0.52 | | | | Binary |
| ir400v2.5 | 6.38 | 10.58 | 3.78 | 0.53 | 6.96 | 92.67 | 0.84 | Triple |
| | 2.99 | 5.76 | 0.84 | 0.03 | 9.58 | 8.13 | 0.66 | Triple |
| ir400v2.6 | 3.61 | 7.43 | 2.39 | 0.55 | | 80.65 | 0.71 | Quadruple |
| | 4.43 | 7.14 | 1.84 | 0.8 | | | | |

**Table S5.**

**Test simulations involving two identical disks ($m_s$0.3$m_d$0.18 model in fig. S1) with mutual disk inclination of *i* degrees.** Encounters with mutual disk inclinations smaller than the disk opening angle, about 5 degrees, can form FFOs.

| Case | Number of BCs | Number of FFOs (binaries) | Sink mass (disk mass) for FFOs |
|---|---|---|---|
| rp300v2.5i3 | 0 | 1 | 17.77 (6.77) |
| rp300v2.5i5 | 1 | 1 | 10.44 (3.64) |
| rp300v2.5i7 | 0 | 0 | |
| rp300v2.5i9 | 0 | 0 | |
| rp400v2.5i3 | 0 | 1 | 15.31 (10.64) |
| rp400v2.5i5 | 0 | 0 | |
| rp400v2.5i7 | 0 | 0 | |
| rp400v2.5i9 | 0 | 0 | |

**Table S6.**

**Low-resolution simulations involving the two barotropic disk models in fig. S1 but at a mass resolution of 0.0002 Jupiter mass.** The results are compared to those of table S2 in fig. S5. In general, the low-resolution simulations form more massive FFOs than the high-resolution simulations.

| Case | Number of BCs | Number of FFOs (binaries) | Sink mass (disk mass) for FFOs |
|---|---|---|---|
| nr300v2.1LR | 2 | 0 | |
| nr300v2.2LR | 0 | 1 | 26 (9.64) |
| nr300v2.3LR | 1 | 0 | |
| nr300v2.4LR | 1 | 1 | 17.35 (4.54) |
| nr300v2.5LR | 0 | 1 | 17.25 (4.98) |
| nr300v2.6LR | 1 | 1 | 15.44 (3.83) |
| nr300v2.7LR | 1 | 1 | 15.95 (5.57) |
| nr300v2.8LR | 0 | 1 | 22.09 (8.41) |
| nr300v2.9LR | 0 | 1 | 17.11 (5.54) |
| nr300v3.0LR | 0 | 1 | 7.77 (3.13) |
| nr300v4.0LR | 0 | 0 | |
| nr400v2.0LR | 1 | 1 | 20.11 (7.57) |
| nr400v2.5LR | 0 | 0 | |
| nr400v3.0LR | 0 | 0 | |

**Movie S1.**

**Animation of Fig. 1**. It shows the formation of a free-floating PMO binary in the rp400v2.65 model of table S1.
https://drive.google.com/file/d/1MO0B9aKcHkrLTIuCEI6pnEbwb89107gs/view?usp=share_link

**Movie S2.**

**Animation of fig. S7**. It shows the formation of two free-floating PMO triples in the ir400v2.5 model of table S3.
**https://drive.google.com/file/d/1UaGZ7yf_nweVYQD71GrmZYLZA5YVXFVA/view?usp=share_link**